\documentclass[aps,prl,twocolumn,showpacs]{revtex4}
\usepackage{epsfig}
\usepackage{times}
\begin{document}

\title{Success-driven distribution of public goods promotes cooperation but preserves defection}

\author{Matja{\v z} Perc}
\affiliation{Faculty of Natural Sciences and Mathematics, University of Maribor, Koro{\v s}ka cesta 160, SI-2000 Maribor, Slovenia}

\begin{abstract}
Established already in the Biblical times, the Matthew effect stands for the fact that in societies rich tend to get richer and the potent even more powerful. Here we investigate a game theoretical model describing the evolution of cooperation on structured populations where the distribution of public goods is driven by the reproductive success of individuals. Phase diagrams reveal that cooperation is promoted irrespective of the uncertainty by strategy adoptions and the type of interaction graph, yet the complete dominance of cooperators is elusive due to the spontaneous emergence of super-persistent defectors that owe their survival to extremely rare microscopic patterns. This indicates that success-driven mechanisms are crucial for effectively harvesting benefits from collective actions, but that they may also account for the observed persistence of maladaptive behavior.
\end{abstract}

\maketitle

The Gospel of St. Matthew states: ``For to all those who have, more will be given.'' Roughly two millennia latter sociologist Robert K. Merton \cite{merton_sci68} was inspired by this writing and coined the term ``Matthew effect'' for explaining discrepancies in recognition received by eminent scientists and unknown researchers for similar work. A few years earlier physicist and information scientist Derek J. de Solla Price \cite{price_sci65} actually observed the same phenomenon when studying the network of citations between scientific papers, only that he used the phrase cumulative advantage for the description. To most physicists, however, preferential attachment will be best known due to the seminal paper by Barab{\'a}si and Albert \cite{barabasi_s99}, who used the concept ingeniously to explain the emergence of scaling in growing random networks. Other common variations of the phrase include rich-get-richer and success-breeds-success \cite{egghe_jasis95}, all implying that initial advantages are often self-amplifying and tend to snowball over time. No matter the wording, be it the accumulation of wealth \cite{smith_04} or citations \cite{redner_pt05}, the making of new friends \cite{tomkins_KDD06}, or the longevity of one's career \cite{petersen_pnas11}, this simple yet fascinatingly powerful phenomenon arguably influences many facets of our existence.

In this paper we investigate how the Matthew effect affects the evolution of cooperation in the public goods game. The public goods game \cite{sigmund_10, nowak_06} is played in groups and captures the essential social dilemma in that collective and individual interests are in dissonance. Players must decide simultaneously whether they wish to contribute to the common pool or not. All the contributions are then multiplied to take into account synergetic effects of cooperation, and the resulting amount is divided equally among all group members irrespective of their initial decision. From the perspective of each individual, defection is clearly the rational decision to make as it yields the highest personal income if compared to other members of the group. However, if nobody decides to invest the group fails to harvest the benefits of a collective investment and the society evolves towards the ``tragedy of the commons'' \cite{hardin_g_s68}. The sustenance of cooperation in sizable groups of unrelated individuals, as is the case by the public goods game, is particularly challenging since group interactions tend to blur the trails of those who defect. Unlike by pairwise interactions, reciprocity \cite{axelrod_84, nowak_n98} often fails as it is not straightforward to determine whom to reciprocate with. Social enforcement, on the other hand, may work well, although it is challenged by the fact that it is costly (see \cite{sigmund_tee07} for a review). Other prominent ways of promoting cooperation in public goods games include the introduction of volunteering \cite{hauert_s02}, social diversity by means of complex interaction networks \cite{santos_n08}, heterogeneous wealth distributions \cite{wang_j_pre10b}, and institutionalized punishment \cite{sigmund_n10}.

Inspired by the seminal works on games on coevolutionary and social networks \cite{abramson_pre01, ebel_pre02, zimmermann_pre04, eguiluz_ajs05, pacheco_prl06}, the most recent advances on this topic \cite{wu_t_pre09, cardillo_njp10, zschaler_njp10, lee_s_prl11, gomez-gardenes_chaos11} (see \cite{perc_bs10} for a review), as well as the seeming omnipresence of the Matthew effect in social interactions, we consider the public goods game where the distribution of multiplied contributions is not equally shared amongst all the group members, but rather it depends on the evolutionary success of each individual. Naturally, the more successful an individual is the higher its share of the public good.

Assuming structured interactions, $L^2$ players are arranged into overlapping groups of size $G$ such that every player is surrounded by its $G-1$ nearest neighbors. Accordingly, each individual belongs to $g=G$ different groups. Initially each player on site $x$ is designated either as a cooperator ($s_x = C$) or defector ($s_x = D$) with equal probability. Cooperators contribute a fixed amount (here considered being equal to $1$ without loss of generality) to the public good while defectors contribute nothing. The sum of all contributions in each group is multiplied by the factor $r>1$ and the resulting public goods are distributed amongst all the group members. If $s_x = C$ the payoff of player $x$ from every group $g$ is $P_{C}^g=M_x r N_{C}^g/G - 1$ and if $s_x = D$ the payoff is $P_{D}^g=M_x r N_{C}^g/G$, where $N_{C}^g$ is the number of cooperators in group $g$ while $M_x$ is the factor by means of which the Matthew effect is introduced. Initially all players have $M_x=1$, and so without further modifications the setup returns the classical spatial public goods game \cite{szolnoki_pre09c}. Here, however, each time player $x$ successfully enforces its strategy on another player $y$, $M_x=M_x+\Delta$ and $M_y=M_y-\Delta$, where $\Delta>0$ is a free parameter. Thus, in the next round player $x$ will receive a higher share of the public goods while player $y$ will receive, to the same extent, a smaller one. Note that since the fact that player $x$ was able to enforce its strategy on player $y$ already implies that the former is more successful, this simple coevolutionary rule will strengthen this further whilst at the same time additionally degrading player $y$, thus concisely introducing the Matthew effect into the public goods game. Importantly, the coevolutionary rule is strategy independent and at no point in time assumes any public goods being lent or not spent, i.e., $L^{-2}\sum_x M_x=1$ at all times.

\begin{figure}
\centerline{\epsfig{file=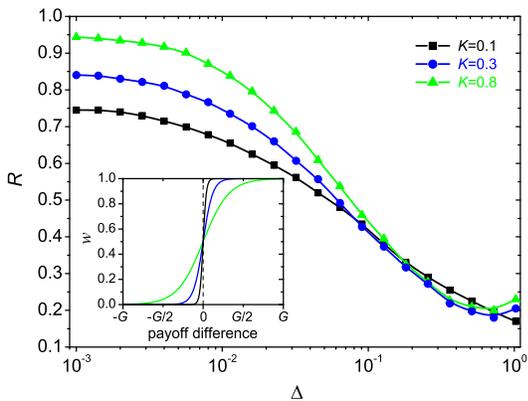,width=7.0cm}}
\caption{(Color online) Critical $R$ at the $D \to C+D$ phase transition in dependence on $\Delta$ for three different values of $K$. Success-driven distribution of public goods strongly decreases the multiplication factor needed for cooperation to survive, irrespective of $K$. Inset depicts the probability of strategy adoption $w$ in dependence on the payoff difference for the three $K$ [same color (gray scale) as in the main panel]. Square lattice with $G=5$ was used as the interaction graph.}
\label{phaseK}
\end{figure}

Monte Carlo simulations are carried out comprising the following elementary steps. A randomly selected player $x$ plays the public goods game with its $G$ partners as a member of all the $g=1, \ldots, G$ groups, whereby its overall payoff is thus $P_{s_x} = \sum_g P_{s_x}^g$. Next, player $x$ chooses one of its nearest neighbors at random, and the chosen co-player $y$ also acquires its payoff $P_{s_y}$ in the same way. Finally, player $x$ enforces its strategy $s_x$ onto player $y$ with a probability $w(s_x \to s_y)=1/\{1+\exp[(P_{s_y}-P_{s_x})/GK]\}$, where $K$ quantifies the uncertainty by strategy adoptions due to errors in decision making or incomplete information (see inset of Fig.~\ref{phaseK}), and $G$ normalizes the effect for different interaction graphs \cite{szolnoki_pre09c}. Each Monte Carlo step (MCS) gives a chance for every player to enforce its strategy onto one of the neighbors once on average. The average frequencies of cooperators ($\rho_{C}$) and defectors ($\rho_{D}$) were determined in the stationary state after sufficiently long relaxation times. Depending on the actual conditions (proximity to phase transition points and the typical size of emerging spatial patterns) the linear system size was varied from $L=200$ to $1600$ and the relaxation time was varied from $10^4$ to $10^7$ MCS to ensure proper accuracy.

\begin{figure}
\centerline{\epsfig{file=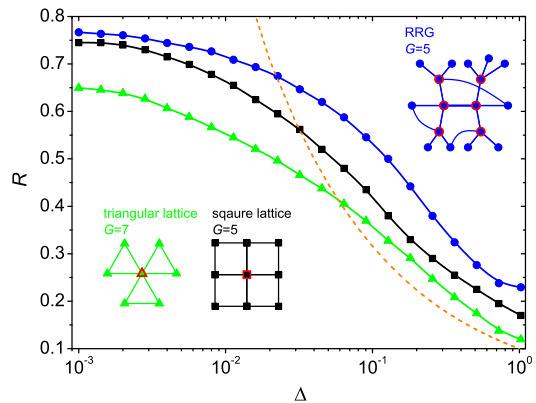,width=7.0cm}}
\caption{(Color online) Critical $R$ at the $D \to C+D$ phase transition in dependence on $\Delta$ for three different interaction graphs. Insets show the graphs schematically in the color (symbol type) corresponding to the results in the main panel, with the group size $G$ indicated for vertices encircled red (gray). As in Fig.~\ref{phaseK} the promotion of cooperation is clearly inferable, irrespective of the properties of the interaction graph. Here $K=0.1$ was used in all three cases. The dashed orange line shows the result as predicted by the mean-field approximation (see main text for details).}
\label{phaseL}
\end{figure}

In Fig.~\ref{phaseK} we plot the critical value of $r$, defined as the minimally required $R=r/G$ (note that the normalization with $G$ enables comparisons with results on graphs other than the square lattice \cite{szolnoki_pre09c}) where $\rho_{C}$ first becomes $>0$, in dependence on $\Delta$. It can be observed that the stronger the Matthew effect, the lower the multiplication factor needed for the sustenance of cooperation. This holds irrespective of $K$. While $\Delta=0.001$ essentially returns the classical version of the game, with the same $R$ as reported in \cite{szolnoki_pre09c}, a sharp descent towards as low as $R=0.17$ follows for larger $\Delta$. The effect saturates and may even revert slightly when $\Delta$ becomes comparable to $G$, which is due to the fact that $M_x$ must not exceed $G$ since \textit{all} the accumulated public goods within a group are then already assigned to player $x$. To test the generality of the observed promotion of cooperation further, we plot in Fig.~\ref{phaseL} the critical $R$ for three different interaction graphs (see insets) with very distinct properties that are know to vitally affect the evolution of cooperation \cite{szabo_pr07}. Namely, a regular graph with zero clustering coefficient (square lattice), a regular graph with a high clustering coefficient (triangular lattice), and a random regular graph having no local structure. Irrespective of these details the Matthew effect promotes cooperation equally well, as can be concluded from the descending $R(\Delta)$ phase transition lines.

An inspection of the stationary distribution of $M_x$ values for intermediate $\Delta$ reveals a double peaked Gaussian with maxima approximately at $-G/2$ and $G/2$ respectively. This is easily traced back to the nature of the Matthew effect in that it segregates the population into successful (peaked around $G/2$) and unsuccessful (peaked around $-G/2$) players. What is more intriguing is the observed promotion of cooperation, which was in the past typically associated with strongly heterogeneous (e.g. power law or exponential) distributions, whereby it was argued that the high-impact (in our case equivalent to high $M_x$) players stabilize cooperation while defectors are doomed by means of a negative feedback effect \cite{szabo_pr07}. This well-known explanation applies only partially in our case. In fact, here cooperators benefit also from the dynamical reshaping of the evolutionary landscape (defined by $M_x$), which renders the defective strategy maladaptive and thus gives way to a mixed $C+D$ phase. This can be confirmed analytically by means of a simple well-mixed approximation of our model. Namely, by plugging into the well-mixed ansatz \cite{hofbauer_98} $\dot{\rho_C}=\rho_C(1-\rho_C)(P_C-P_D)$ the payoffs as defined above, treating $N_C/G$ as $\rho_C$ and introducing Gaussian distributed values $\xi$ with standard deviation $\sigma$ for the difference in $M$ (as motivated by the observed distribution of $M_x$ values), we obtain $\dot{\rho_C}=\rho_C(1-\rho_C)(r \rho_C \xi -1)$. Following a first-order small-noise expansion \cite{horsthemke_84} we finally have $\dot{\rho_C}=\rho_C(1-\rho_C)(\sigma^2 r \rho_C(\rho_C-3\rho_C^2/2)-1)$. From $\dot{\rho_C}=0$ and the sign of the second derivative we find $\rho_C=0$ and $\rho_C=1$ as unstable steady states. The new stable steady state of $\rho_C$ comes from $\sigma^2 r \rho_C(\rho_C-3\rho_C^2/2)=1$ (the polynomial has three roots, two of which complex conjugates), which however is to cumbersome to be expressed here explicitly. What is relevant is that one obtains $r \propto \sigma^{-0.5}$ for the $D \to C+D$ phase transition, which can be verified easily by straightforward numerical integration of $\dot{\rho_C}$. Since we lost $G$ in the well-mixed approximation $r$ may be rescaled to account for $R$ as obtained from Monte Carlo simulations and $\sigma$ takes on the role of $\Delta$. Preserving the slope of the dependence this yields the dashed orange line depicted in Fig.~\ref{phaseL}. While it is by no means implied that this approximation gives a good fit to the Monte Carlo simulations, it nevertheless confirms the effect by means of a simple analytically treatable model.

\begin{figure}
\centerline{\epsfig{file=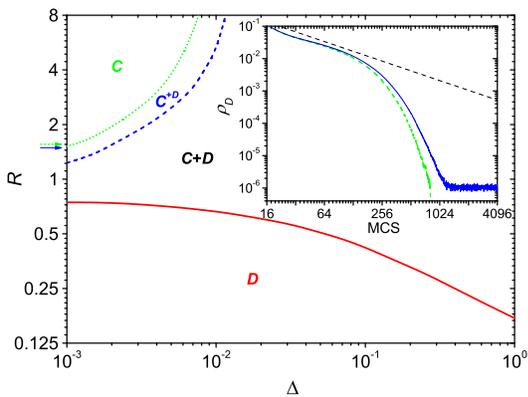,width=7.0cm}}
\caption{(Color online) Strategy regions in the $\Delta-R$ parameter space for the square lattice at $K=0.1$. While $R$ needed for cooperation to survive decreases steadily with increasing $\Delta$ (solid red line), the opposite holds for the extinction of defectors. The transition to the pure $C$ region (dotted green line) is preceded by the emergence of a peculiar $C^{+D}$ region (dashed blue line), where a minute fraction of defectors is able to survive in the presence of dominating cooperators. The inset shows two characteristic time courses for just above [dashed green line (top arrow at $R=1.5$)] and below [solid blue line (bottom arrow at $R=1.47$)] the $C^{+D} \to C$ transition line for $\Delta=0.001$. Within the $C^{+D}$ region defectors utilize their one-in-a-million survival chance to evade extinction. Although in the inset only the first 4096 MCS are depicted, we have verified that at $R=1.47$ $\rho_D>0$ for up to $10^7$ MCS at $L=1600$ system size. The dashed black line in the inset is proportional to MCS$^{-1}$.}
\label{phaseF}
\end{figure}

\begin{figure}
\centerline{\epsfig{file=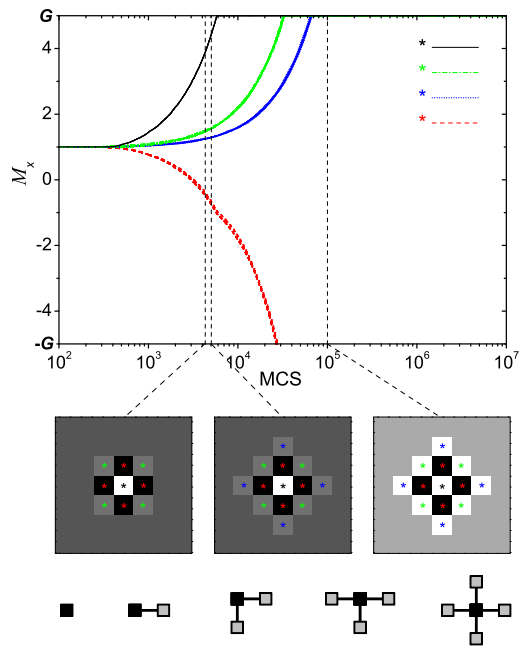,width=7.0cm}}
\caption{(Color online) Uppermost panel shows time courses of $M_x$ for the ``super-persistent'' defector (solid black), its four nearest neighbors (dashed red), its four next-nearest neighbors (dash-dotted green) and the four next-next-nearest neighbors (dotted blue). The three middle panels depict gray-scale coded (white maximal, black minimal) values of $M_x$ around the ``super-persistent'' defector at times corresponding to the vertical dashed lines in the uppermost panel. Colored stars serve the identification of the neighbors corresponding to the colors (shades of gray) used in the uppermost panel. The bottommost row depicts possible configurations of defectors (gray) around the ``super-persistent'' defector (black). Results were obtained on a square lattice of linear size $L=1600$ at $K=0.1$, $R=1.47$ and $\Delta=0.001$.}
\label{persist}
\end{figure}

The promotion of cooperation, however, is only one observation following from the introduction of the Matthew effect. There is another, namely the preservation of defectors to be demonstrated below, which is rooted in extremely rare small-region effects originating from the spatiality of the considered model. Figure~\ref{phaseF} depicts the strategy regions in the $\Delta-R$ parameter space for the square lattice, from where the peculiar behavior can be inferred. In particular, as the critical $R$ inducing $D \to C+D$ falls, the critical $R$ required for $C+D \to C$ raises sharply with increasing $\Delta$. Extensive Monte Carlo simulations using $L=1600$ reveal a narrow intermediate region denoted as $C^{+D}$, where the most minute fraction of defectors can prevail indefinitely in the sea of cooperators. The inset features two time courses just above and below the $C^{+D} \to C$ transition line (see arrows in the main panel). What at first appears to be algebraically slow relaxation (marked by the dashed line in the inset) is facilitated by the coevolutionary impact of the Matthew effect to become either an absorbing $C$ region (green) or the $C^{+D}$ region (blue), where approximately one defector is able to survive amongst $10^6$ cooperators (note that the temporal courses are averages over 10 independent realizations). We also note that we use region rather than phase for the $C+D \to C^{+D} \to C$ transitions since we were unable to completely resolve the large system size limit.

Zooming in on the neighborhoods around these ``super-persistent'' defectors gives vital clues with regards to their survival. Figure~\ref{persist} shows the time evolution of $M_x$ for one such defector and its neighbors. Evidently, it is possible, although highly unlikely, that isolated defectors become fully protected by their neighbors by means of the spontaneous emergence of successive microscopic hierarchies (expressed in terms of $M_x$) in space. The three middle panels show that while the central player (defector) enjoys the benefits of the highest possible $M_x$, its four nearest neighbors have the lowest possible $M_x$ associated to them. Further crucial is then the fact that all the neighbors of the four nearest neighbors, like the central defector, also all eventually acquire the highest possible $M_x$. This makes it impossible for the four potential donors of the new strategy to the central defector to overtake it. Even though they may occasionally succeed in adopting the cooperative strategy, it is impossible to enforce the latter onto the central defector because they are at an inherent disadvantage due to the unfavorable outcome of the Matthew effect in their immediate neighborhood. Consequently, the only possible configurations (showing defectors only) around the ``super-persistent'' defector are those depicted in the bottom row of Fig.~\ref{persist}, with zero chances of full cooperator dominance. It is to be emphasized that although one defector amongst a million cooperators seems negligible, from the evolutionary viewpoint it nevertheless preserves the seed for defection to take on a more dominant role should the conditions ever become more favorable. From the view point of statistical mechanics, it is fascinating to learn that coevolutionary rules may, due to the interplay of spatial structure and the dynamics of the coevolving quantity, induce spatially extremely localized microscopic patterns that prevent phase transitions into absorbing states despite the fact that the strategy destined to die out is completely maladaptive (note that in the absence of the Matthew effect cooperators reach full dominance on the square lattice at $R \approx 1.1$ \cite{szolnoki_pre09c}).

In sum, we have demonstrated that a simple strategy-independent coevolutionary rule mimicking the Matthew effect promotes cooperation in group interactions amongst unrelated individuals. This can be partially attributed to the emergent Gaussian distribution of factors determining the distribution of public goods, but also to the inability of defectors (or their lesser ability if compared to cooperators) to adapt to continuously changing evolutionary landscapes, as demonstrated by an analytically treatable well-mixed approximation. However, the Matthew effect may also give rise to spatially highly localized microscopic patterns that protect maladaptive strategies despite of their obvious evolutionary disadvantage, thus preventing absorbing phases and preserving an option of a comeback of seemingly exterminated traits. We hope that this work will inspire further research in the yet unexplored directions concerning the impact of coevolutionary rules on the outcome of games governed by group interactions.

This research was supported by the Slovenian Research Agency (grants Z1-2032 and J1-4055).

\end{document}